\documentstyle[12pt,aaspp4,tighten]{article}
\begin{document}

\newcommand{\msun}{M{$_{\odot}$}} 
\newcommand{\lsun}{L{$_{\odot}$}}
\newcommand{\rsun}{R{$_{\odot}$}}
\newcommand{\lsim}{\;\lower.6ex\hbox{$\sim$}\kern-7.75pt\raise.65ex\hbox{$<$}\;}
\newcommand{\gsim}{\;\lower.6ex\hbox{$\sim$}\kern-7.75pt\raise.65ex\hbox{$>$}\;}

\title{Star formation in clusters: a survey of compact mm-wave sources
in the Serpens core}

\author{Leonardo Testi and Anneila I. Sargent}

\affil{Division of Physics, Mathematics and Astronomy, California 
Institute of Technology, MS~105-24, Pasadena, CA~91125
(lt@astro.caltech.edu,afs@astro.caltech.edu)}

\begin{abstract}

We report the results of a millimeter interferometric survey of compact
3~mm continuum sources in the inner $5.5^\prime\times 5.5^\prime$ region
of the Serpens core. We detect 32 discrete sources above $4.0$~mJy/beam,
21 of which are new detections at millimeter wavelengths. By comparing
our data with published infrared surveys, we estimate that 26 sources are
probably protostellar condensations and derive their mass assuming optically
thin thermal emission from dust grains. The mass spectrum of the clumps,
dN/dM$\sim$M$^{-2.1}$, is consistent with the stellar initial mass function,
supporting the idea that the stellar masses in young clusters are determined
by the fragmentation of turbulent cloud cores.

\end{abstract}

\keywords{ ISM: clouds -- ISM: radio continuum -- stars: formation }

\section{Introduction}
\label{sintro}

The theory of isolated star formation is in general fairly well understood
(e.g. Shu et al.~\cite{SAL87}). One notable flaw, however, is its failure
to predict the resulting stellar masses. Stars in the field are known
to be distributed according to a well defined mass function
(e.g. Salpeter~\cite{S55}; Kroupa, Tout \& Gilmore~\cite{KTG93}),
and a complete theory of star formation must be able to predict
this mass distribution, the Initial Mass Function (IMF). On the other hand,
it seems very likely that most stars form in clusters rather
than in isolation (Lada~\cite{L92};
Zinnecker, McCaughrean \& Wilking~\cite{ZMW93};
Lada, Alves \& Lada~\cite{LAL96}; Testi, Palla \& Natta~\cite{TPN98}).
Thus any understanding of the star formation process must also be closely
related to the way in which the IMF originates in clusters.
Indeed, young embedded clusters appear to be populated by stars with a 
mass distribution very close to that observed in the solar neighbourhood
(Hillenbrand~\cite{H97}; Lada et al.~\cite{LAL96}).

The IMF may result naturally from the protostellar accretion process. 
Adams \& Fatuzzo~(\cite{AF96}), for example, suggest outflow/inflow
interactions as the mechanism that determines stellar masses. 
Alternatively, the stellar mass distribution may be caused by the 
fragmentation process in turbulent, cluster-forming, dense cores
(Myers~\cite{M98}; Klessen, Burkert \& Bate~\cite{KBB98}).
If the fragmentation hypothesis is correct, the
prestellar condensations in cluster-forming molecular cores should be very
close to the IMF of the stars in young embedded clusters and in the field.
Interferometric molecular line studies of cloud cores 
(Pratap, Batrla \& Snyder~\cite{PBS90}; Kitamura, Kawabe \&
Ishiguro~\cite{KKI92}) yield contradictory results, depending on the 
chemical and physical conditions assumed for the emitting gas.
Observational evidence supporting the fragmentation origin of the IMF
has been obtained by Motte, Andr\'e \& Neri~(\cite{MAN98}), in 1.3~mm
continuum maps of $\rho$-Ophiuchi, a cluster forming core,
using the IRAM 30-m telescope. The mass spectrum of the prestellar
and protostellar clumps within the core appear consistent with the IMF.
Surveys of additional cluster-forming molecular cores are needed
to confirm that this result holds generally and to constrain the theoretical
models. 

At the angular resolution achievable with the IRAM-30m/bolometer system
at 1.3~mm, $\sim 13^{\prime\prime}$, it is possible to probe detailed
cloud core structure only in the star forming regions closest to the Sun,
such as Taurus or Ophiuchus.
Millimeter wavelength interferometers, however, offer both high spatial
resolution and high sensitivity, and at these wavelengths the dust emission is
probably optically thin, allowing a reasonable estimate of clump 
masses. Moreover, thanks to the interferometric filtering capability, smooth,
extended emission from the molecular cloud core in which clumps are embedded 
is resolved out. Using the Owens Valley Radio Observatory (OVRO)
millimeter array, we have begun a program of high resolution, millimeter-wave
mapping of molecular cloud cores with a view to establishing whether the
prestellar clump mass function and the IMF are in general similar.
In addition to the above-mentioned advantages, the OVRO array enables
simultaneous observation of molecular line and broad band continuum emission.
Any contamination of the continuum flux by molecular line radiation can
therefore be eliminated, allowing more accurate mass estimates. The 
results of this program should constrain theoretical models of the IMF.

Here we present our results for the Serpens star-forming core (Loren et
al.~\cite{Lea79}; Ungerechts \& G\"usten~\cite{HG84}; White et
al.~\cite{Wea95}). At a distance of 310~pc (de~Lara et al.~\cite{dLea91}), and
with an angular extent of few arcmin, this is an ideal target to search for
compact prestellar and protostellar condensations.  Inside the 1500~M$_\odot$
core is a young stellar cluster of approximate mass 15-40$\rm\,\,M_\odot$
(Strom, Grasdalen \& Strom~\cite{SGS76}; Eiroa \& Casali~\cite{EC92};
Giovannetti et al.~\cite{Gea98}). In addition, far-infrared and submillimeter
observations reveal the presence of a new generation of embedded objects
(Casali et al.~\cite{CED93}, hereafter CED; Hurt \& Barsony~\cite{HB96}).
The improved sensitivity and resolution over the large area, 
$\sim 5.5^\prime\times 5.5^\prime$, covered by our survey have
enabled us to detect a large number of new mm sources and to derive the mass
spectrum of the dust condensations.

\section{Observations and results}
\label{obser}

We observed the $\sim 5.5^\prime\times 5.5^\prime$ inner region of the Serpens
molecular core with the OVRO
millimeter wave array during the period September 1997 -- February
1998. The final map resulted from 50 separate pointings of the telescope
during each transit.
The primary beam size at 99~GHz is $\sim 73^{\prime\prime}$ (FWHM) and 
pointing centers were spaced by $\sim 42^{\prime\prime}$.
Four configurations of the six 10.4-m dishes provided baselines in the range
$\sim 5-80~\rm k\lambda$. This ($u,v$) sampling ensures good sensitivity up
to spatial scales of $\sim 30^{\prime\prime}$ equivalent to 0.045~pc, or
9300~AU, at the distance of Serpens.
All telescopes are equipped with cryogenically cooled SIS receivers
which provided average system temperatures of $\sim 350$~K (SSB) at
the observing frequency.
Continuum observations centered at 99~GHz were made in both (USB and LSB) 
1~GHz wide bands of the analog correlator.
An 8~MHz wide band at 0.125~MHz resolution of the digital correlator
was centered on the CS(2--1) transition at 97.981~GHz at the Serpens core
velocity, $v_{LSR}=8.0$~km/s. Gain and phase were calibrated through
frequent observations of the quasar 1741$-$038. 3C273 and/or 3C454.3 were
used for passband calibration. The flux density scale was determined by
observing Neptune and Uranus and the estimated uncertainty is less
than 20\%. All calibration and editing of the raw data have been performed
with the MMA software package (Scoville et al.~\cite{Sea93}).
The NRAO--AIPS package was used for mapping and analysis.

To produce the final mosaic image of the region, we applied the AIPS VTESS
task, which performs a simultaneous Maximum Entropy (MEM) deconvolution
of all the observed pointings. We used natural weighting and applied 
gaussian tapering to the ($u,v$) data before mapping/deconvolution.
Nevertheless, due to residual phase errors and strong
dirty beam sidelobes, it was necessary to remove the 
three brightest point sources (SMM1, SMM3 and SMM4) before deconvolution.
The final hybrid image of the observed region, shown in Fig.~\ref{fcontmos},
was then obtained by restoring to the MEM mosaic the three clean images
(obtained with the task IMAGR) of the bright sources.
The synthesized beam is $5\farcs5\times
4\farcs3$ FWHM, corresponding to linear resolution $\sim 0.0075$~pc, or
1500~AU, and the noise level is $\sim 0.9\rm\,\, mJy/beam$.
The CS(2--1) data was used to produce a map of the integrated line
emission and we verified that line contamination in our continuum map
is negligible. Detailed analysis of the CS data goes beyond the scope
of the present paper and will be presented separately.

\section{Analysis}
\label{sres}

The angular resolution of the 99~GHz hybrid mosaic of the Serpens core region 
shown in Fig.~\ref{fcontmos}
is a factor of 4 higher than that of the CED sub-millimeter map
of the same region. The sensitivity is more than 10 times 
higher than in the less extensive 3~mm continuum map of McMullin et
al.~(\cite{MMea94}).
The known far-infrared and sub-millimeter
young stellar objects (CED; Hurt \&
Barsony~\cite{HB96}) are indicated by crosses, although their
positional uncertainties are rather large.
A particularly notable case is S68N where the old and new positions 
differ by $\sim 10^{\prime\prime}$,
a discrepancy already pointed out by Wolf-Chase et al.~(\cite{WCea98}).
All these objects are detected in our observations.
In addition, we identified 21 new discrete sources with peak flux densities
above the $\sim 4.5\sigma$ level of $4.0~\rm mJy/beam$. We are confident that
this is an appropriate threshold since all our detected sources
in the S68N/SMM1 region are ``visible'', but at lower confidence level,
in the lower resolution SCUBA
observations at 450~$\mu$m (Wolf-Chase et al.~\cite{WCea98}) and in 
BIMA/IRAM-30m observations at 3 and 1.3~mm (Williams et al. in preparation).

\begin{figure*}
\epsscale{1.0}
\plotfiddle{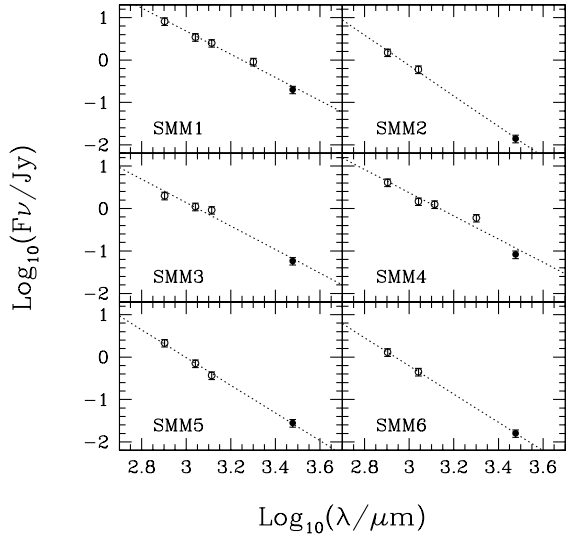}{7cm}{0}{40}{40}{-270}{-60}
\caption[fsmmspec.ps]{\label{fsmmspec} Continuum spectra for the 6 sources 
in common with CED; open circles: their JCMT measurements; filled circles:
OVRO 3~mm fluxes. For each source, a power law spectrum has been fitted
assuming F$_\nu\sim\nu^{\beta+2}$.}
\end{figure*}

We calculated the mass of each of the sources of emission
assuming optically thin dust thermal emission, with
M$_d=S_\nu\,D^2/(\kappa_\nu\,B_\nu(T))$, where $S_\nu$ is the observed
integrated flux density, $D$ is the distance from the Sun, $B_\nu(T)$ is the
Planck function at the assumed dust temperature $T$, and $\kappa_\nu$ is the 
dust mass opacity coefficient. We assumed $T=15$~K and $\kappa_\nu=
\kappa_{230GHz}(\nu/230GHz)^\beta$.
Following Preibish et al.~(\cite{Pea93}),
we adopted $\kappa_{230GHz}=0.005\rm\,\,cm^2\,g^{-1}$, assuming a gas
to dust ratio of 100 by mass. This value of $\kappa_{230GHz}$ is consistent
with recent measurement in the cold cloud core IC~5146 (Kramer et
al.~\cite{KALLUW98}), and agrees to within a factor of two with that 
derived for the envelopes around young stellar objects,
$\sim 0.01\rm\,\,cm^2\,g^{-1}$ by Ossenkopf \& Henning~(\cite{OH94}).
We derived approximate values for $\beta$ by fitting a power law
of the form F$_\nu\sim\nu^{\beta+2}$ to our 3~mm flux and the sub-mm 
measurements of CED, for each of our six common sources.
These fits are shown as dotted lines in Fig.~\ref{fsmmspec}.
In general the OVRO point is consistent with the single dish
fluxes, indicating that the sources are compact and that there is no
significant missing flux from spatially extended envelopes.
Values of $\beta$ range from 0.7 to 1.6,
and we adopted a mean of 1.1 for all the sources in our map.
Our detection limit of $4.0$~mJy/beam then
corrensponds to a mass $\sim 0.4$~\msun\ and, assuming a mean molecular
weight of $2.33$, a beam-averaged  H$_2$ column density of
$\sim 3\times 10^{23}\rm\,\, cm^{-2}$. The mass of each of the clumps
and our mass detection limit depend on the adopted parameter values.
Nevertheless, as long as the emission is optically thin and
all clumps have similar temperatures and opacities,
the {\it shape} of the mass function is independent of the particular
values assumed.

We assume that the mm-sources are a collection of prestellar
condensations, collapsing protostars or
circumstellar structures around more evolved objects, all of which are
expected to show significant continuum emission at mm wavelengths
(e.g. Shu et al.~\cite{SAL87}; Andr\'e~\cite{A96}).
A spherical, isothermal gas cloud with a radius of 2000~AU, a temperature
of 15~K and supported by thermal pressure is gravitationally bound
if it is more massive than $\sim 0.3\rm\,\,M_\odot$ (e.g. Bonnor~\cite{B56}).
Given that 2000~AU is the upper radius for detected sources and $\sim
0.4\rm\,\,M_\odot$ the lower mass limit, all the clumps are likely to be bound.
To ensure that the mass function for condensations (clumps) that will
eventually produce stars is not contaminated by young stellar objects,
we checked for counterparts to our 3~mm sources in the near infrared (NIR)
observations of Giovannetti et al.~(\cite{Gea98}) and the 12~$\mu$m
detections by Hurt \& Barsony~(\cite{HB96}). Only two 12~$\mu$m sources,
PS1 and PS2, and 4 additional NIR sources (which include the two NIR sources
associated with SMM5 and SMM6) could be considered possible
associations; we discarded these 6 sources from our list.
According to the arguments presented by Andr\'e~(\cite{A96}), the
lack of infrared detections suggests that all other clumps, $\sim 80\%$
of the total, are either prestellar or protostellar in nature.
This is a much larger fraction than found in $\rho$-Oph 
by Motte et al.~(\cite{MAN98}); their higher mass sensitivity limit
enables an increased detection rate for faint
circumstellar material around young stars.

\begin{figure*}
\epsscale{1.0}
\plotfiddle{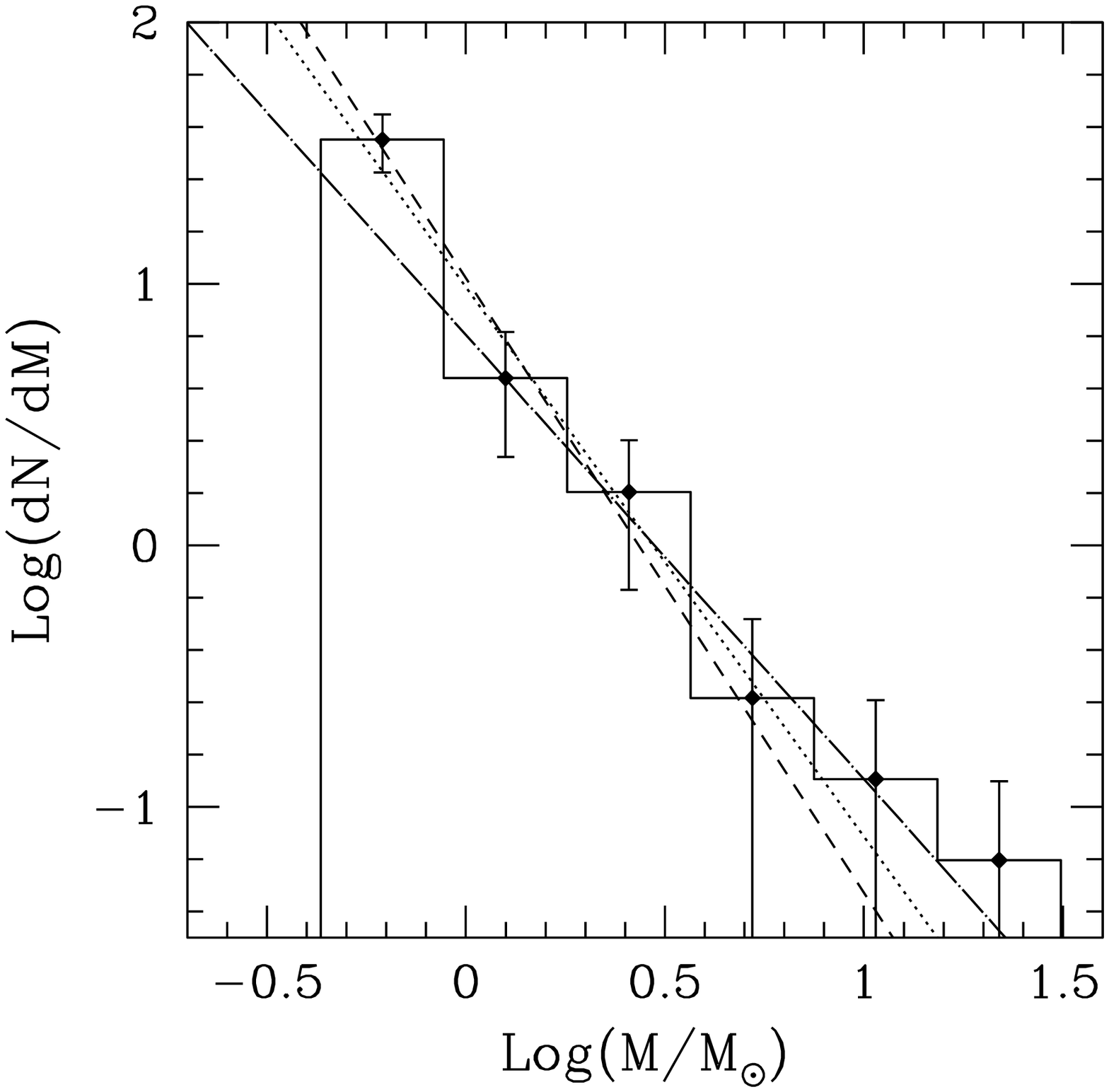}{6cm}{0}{40}{40}{-250}{-80}
\plotfiddle{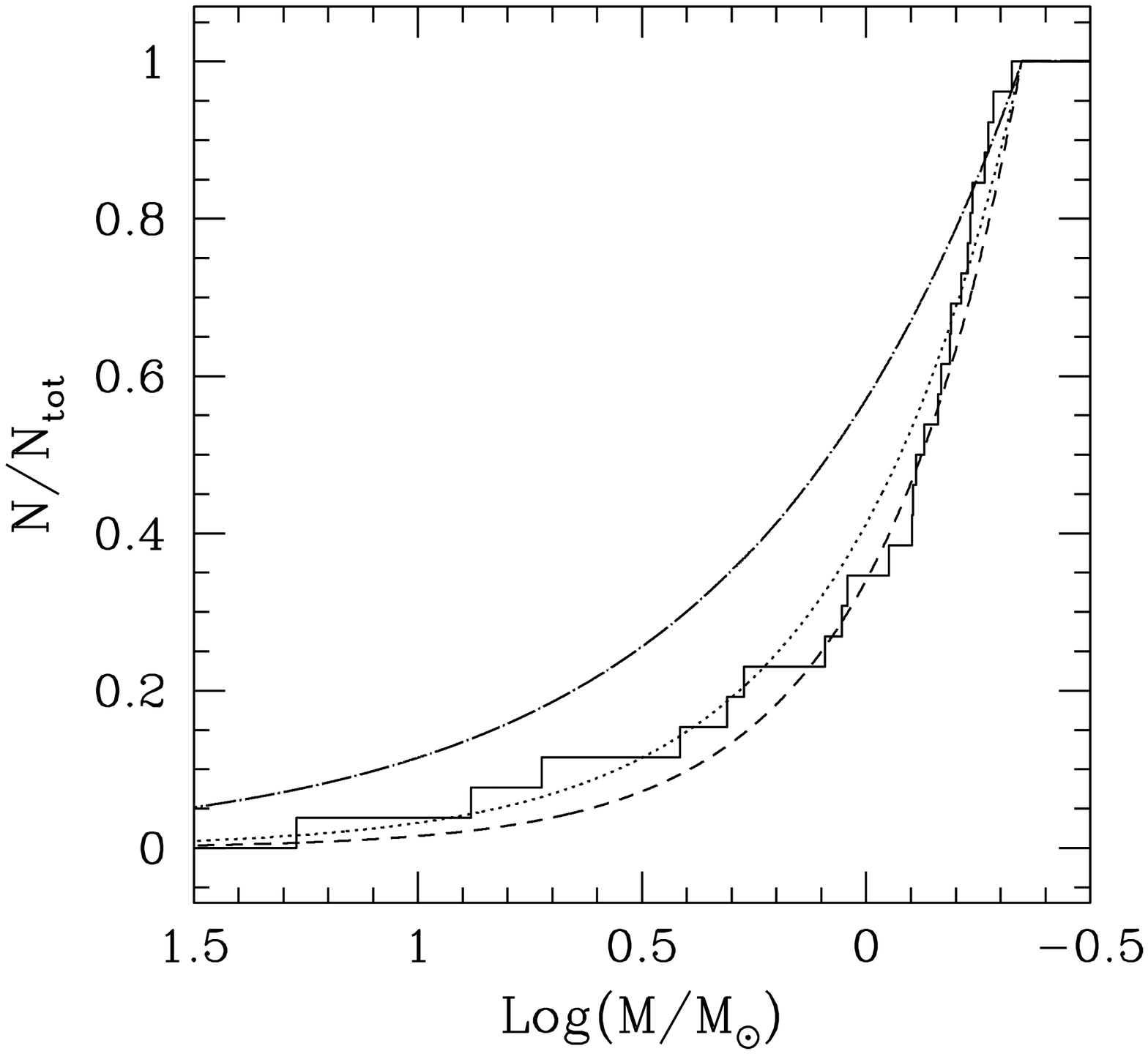}{0cm}{0}{40}{40}{-25}{-58}
\caption[fmspec.ps]{\label{fmspec} Left: the mass spectrum for the 3mm
continuum sources.
The dotted line is the best fitting power law, dN/dM$\sim$M$^{-2.1}$;
the dashed line represents the Salpeter IMF, dN/dM$\sim$M$^{-2.35}$;
the dot-dashed line is a $-$1.7 power--law.
Right: the normalized cumulative mass distribution.
Dotted, dashed and dot-dashed lines are as on the left panel.}
\end{figure*}

In Fig.~\ref{fmspec} the mass spectrum and the cumulative mass function
of the remaining mm sources are shown. The best fitting power law,
$\rm dN/dM\sim M^{-2.1}$, is represented by a dotted line. 
That corresponding to the Salpeter~(\cite{S55})
local IMF, $\rm dN/dM\sim M^{-2.35}$, is a dashed line
while the $-$1.7 power--law spectrum of gaseous clumps is a dot-dashed
line (Kramer et al.~\cite{Kea98}; Williams, Blitz \& McKee~\cite{Wea98}).
Our data are too scant to be compared with the field stars
IMF derived by Kroupa et al.~(\cite{KTG93}). Nevertheless,
our fitted power law index is very close to their $a_2=2.2$ index
for solar and slightly subsolar masses and we view
the agreement between the observed mass spectrum for the clumps and
the stellar mass function for the field stars as very promising.
The cumulative mass function for the clumps, which does not rely on data
binning, provides a more robust comparison; the $-$1.7 power--law
is rejected by the Kolmogorov-Smirnov test at the 98\% confidence
level.

\section{Implications}
\label{sdisc}

\begin{figure*}
\epsscale{1.0}
\plotfiddle{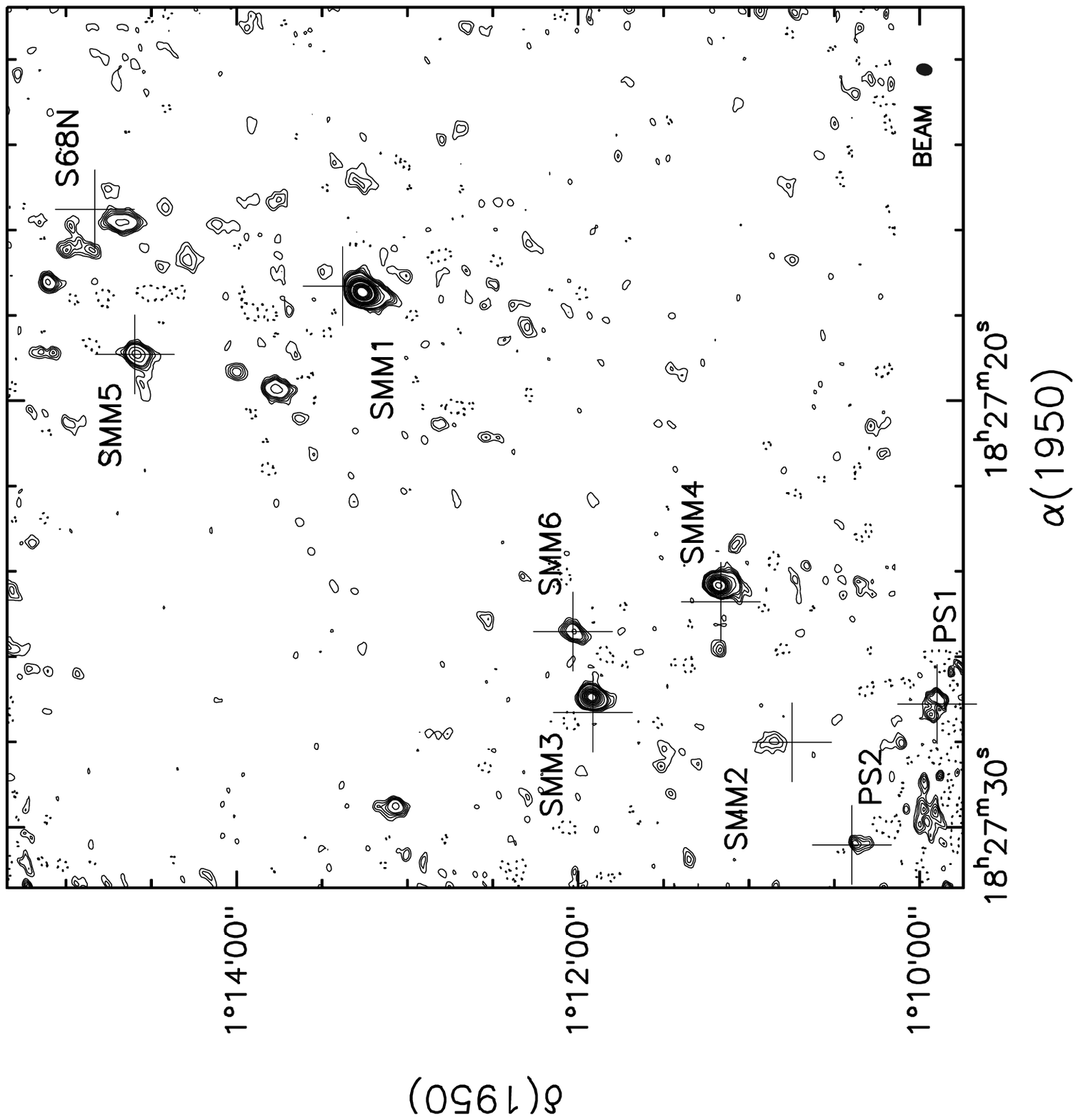}{13.0cm}{270}{90}{90}{-250}{480}
\caption[fcontmos.ps]{\label{fcontmos} OVRO 3~mm continuum mosaic
of the Serpens core. Contour levels are $-$2.7, 2.7 to 6.3 by 0.9,
10 to 42 by 4, 55 to 105 by 10~mJy/beam.
The positions of the known sub-millimeter sources 
(CED) and far-infrared sources (Hurt \& Barsony~\cite{HB96})
are marked by crosses. Note that we detect all the sources already 
identified and have refined the positional accuracy. In addition, numerous
new sources can be seen.
The synthesized beam, $5\farcs5\times4\farcs3$ (FWHM),
is shown as a filled ellipse in the lower right corner.}
\end{figure*}

Our results indicate that the mass spectrum of the protostellar dust
condensations in the Serpens core closely resembles the local IMF. A 
similar result was found for condensations in the $\rho$~Oph cloud
core (Motte et al.~\cite{MAN98}). With $\rm dN/dM\sim M^{-2.1}$,
the slope of the clump mass spectrum is substantially steeper than that 
derived for gaseous clumps.
This strongly suggests that the stellar IMF results from
the fragmentation process in turbulent cloud cores, rather than from 
stellar accretion mechanisms. In fact, the observed 3~mm continuum clumps
may be the direct descendants of the ``kernels'' discussed
by Myers~(\cite{M98}) that have evolved through gravitational contraction.
In Fig.~\ref{fcontmos}, the typical distance between discrete sources 
is $\sim 0.03$--$0.06\rm\,\,pc$, the typical
size of Myers' kernels.

Our conclusions are, of course, preliminary. More definitive statements
must await the completion of high resolution surveys of a substantial
number of cores. The remarkable agreement between the clump mass 
function and the stellar IMF must also be viewed with caution.
A small fraction of our surveyed area has not been observed in the 
near infrared survey of Giovannetti et al.~(\cite{Gea98}), and high resolution,
high sensitivity measurements of the whole region in the mid- to far-infrared
are critical to identifying and eliminating all young stellar objects from our
sample. Nevertheless, our results show that high resolution, millimeter
wave observations of relatively large star forming areas coupled with
near- and mid-infrared surveys, provide an excellent way to 
constrain the origin of the IMF in cluster-forming cloud cores.

\noindent
{\bf Acknowledgements:}  
We thank Francesco Palla, Jonathan Williams, John Carpenter and an
anonymous referee for useful discussions and constructive criticism.
The Owens Valley millimeter-wave array is supported by NSF grant AST-96-13717.
Funding from the C.N.R.--N.A.T.O. Advanced Fellowship program and from
NASA's {\it Origins of Solar Systems} program (through grant
NAGW--4030) is gratefully acknowledged. 
Research at Owens Valley on the formation of young stars and planets
is also supported by the {\it Norris Planetary Origins Project}.

\clearpage

\clearpage

\clearpage

\end{document}